\documentclass[11pt]{article}
\usepackage{PRIMEarxiv}
\usepackage[utf8]{inputenc}
\usepackage{amsmath,amssymb}
\usepackage{booktabs}
\usepackage{multirow}
\usepackage{array}
\usepackage[table]{xcolor}
\usepackage{graphicx}
\usepackage{float}
\usepackage[hidelinks]{hyperref}
\usepackage{placeins}
\usepackage{flafter}

\graphicspath{{figures/}}
\hypersetup{
  pdftitle={Nuisance-Aware Muon Tomography},
  pdfauthor={Zhizheng Zhao, Changhao Qin, Rongfeng Zhang, Zibo Qin, Ziyu Xu, Xingyu Xiao, Qiang Li, and Qite Li},
  pdfkeywords={cosmic-ray muon imaging; muon scattering tomography; material imaging; residual likelihood; open-field calibration; momentum marginalization}}
\renewenvironment{abstract}
  {\centerline{\large\bfseries\scshape Abstract}\par\addvspace{\medskipamount}}
  {\par\addvspace{\medskipamount}}

\newcommand{\Es}{E_{\mathrm{s}}}
\newcommand{\dd}{\mathrm{d}}
\newcommand{\sigpos}{\sigma_{\mathrm{pos}}}
\providecommand{\best}[1]{\textbf{#1}}

\title{\textbf{Nuisance-Aware Muon Tomography}}
\author{
{\bfseries Zhizheng Zhao\textsuperscript{1,\dag}, Changhao Qin\textsuperscript{2},
Rongfeng Zhang\textsuperscript{1}, Zibo Qin\textsuperscript{1}},
\\[-1pt]
{\bfseries Ziyu Xu\textsuperscript{1}, Xingyu Xiao\textsuperscript{1},
Qiang Li\textsuperscript{1}, Qite Li\textsuperscript{1,\dag}} \\[4pt]
\small\textsuperscript{1}Peking University, Beijing 100871, China \\[2pt]
\small\textsuperscript{2}Fudan University, Shanghai 200438, China \\[3pt]
\small\textsuperscript{\dag}Corresponding authors: \href{mailto:zhizhengzhao@outlook.com}{\texttt{zhizhengzhao@outlook.com}}, \href{mailto:liqt@pku.edu.cn}{\texttt{liqt@pku.edu.cn}} \\[3pt]
\small Code and results: \href{https://github.com/zhizhengzhao/NAMT}{https://github.com/zhizhengzhao/NAMT}
}
\date{\today}

\begin{document}
\maketitle

\begin{abstract}
Cosmic-ray muon scattering tomography can image dense, shielded, or
inaccessible objects without an artificial radiation source. In a compact
magnet-free tracker, however, each accepted muon provides only a few hit
positions and no event-by-event momentum measurement. The downstream hit
residual is therefore a compound observable: target scattering, muon momentum,
detector resolution, support material, air scattering, and track extrapolation
all enter the same measured displacement. We introduce Nuisance-Aware Muon
Tomography (NAMT), a residual-likelihood reconstruction method for
magnet-free trackers. The upstream hits define the incident track, downstream
hit residuals carry the scattering signal, and a radiation-length density field
$\lambda=1/X_0$ predicts their material-induced variance through a path
integral. NAMT marginalizes the unmeasured momentum with a shared event-level
scattering scale and uses open-field blank scans to fix detector and
environmental residuals before object reconstruction. On eight Geant4
benchmark scenes spanning strong, weak, and negative scattering contrast,
NAMT-4P reaches a mean area under the ROC curve (AUC) of $0.916$ at $120$k
effective muons and $1$~mm hit error, compared with $0.784$ for ASR, $0.749$
for MLS-EM, and $0.643$ for PoCA. NAMT-3P uses one downstream hit plane in
reconstruction and still reaches $0.909$ mean AUC at the reference setting,
while giving the highest reference mean contrast-to-noise ratio (CNR) and the
best mean AUC at $30$k muons.

\medskip
\noindent\textbf{\emph{Keywords}}\enspace cosmic-ray muon imaging $\cdot$ muon scattering tomography $\cdot$ residual likelihood $\cdot$ open-field calibration $\cdot$ momentum marginalization
\end{abstract}

\section{Introduction}\label{sec:intro}
Cosmic-ray muons are a passive probe for large or shielded objects. They have
been used in geology, archaeology, cargo inspection, and spent-fuel
monitoring~\cite{tanaka2007,alvarez1970,morishima2017,borozdin2003,priedhorsky2003,morris2008,jonkmans2013}.
Recent reviews summarize muography applications, detector technologies,
reconstruction algorithms, and physical
principles~\cite{procureur2018,bonechi2020,bonomi2020,luo2025_muography_review}.
For tracker-based material imaging, the key contrast is multiple Coulomb
scattering, whose variance is governed by the radiation-length density
$\lambda=1/X_0$~\cite{moliere1948,highland1975,lynchdahl1991,pdg2024}.

The native measurement is neither a voxel value nor a sinogram sample. A
tomograph records a short sequence of detector hits, and the material field is
recovered through a stochastic transport model. PoCA and ASR reduce the hit
record to geometric or angle-statistical evidence, whereas MLS-EM remains a
model-based statistical reconstruction built on derived angle--displacement
observations~\cite{borozdin2003,schultz2004,schultz2007,stapleton2014}. These
choices set the observation layer at which nuisance effects enter the inverse
problem. In a magnet-free tracker, the leading nuisance is the per-muon
momentum: scattering variance scales as $(\Es/\beta p c)^2$, where
$\Es\simeq13.6$~MeV is the multiple-scattering constant. Thus a large residual
may mean a denser material, a softer muon, or both. Hit resolution, support
material, air scattering, alignment, and event selection add residual structure
on top of that momentum ambiguity.

\begin{figure}[!htbp]
\centering
\includegraphics[width=\linewidth]{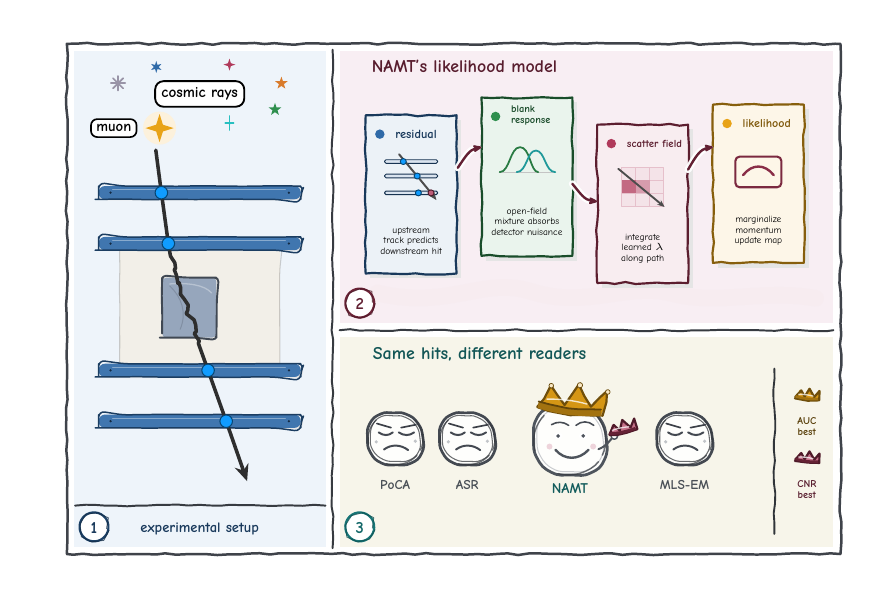}
\caption{\textbf{Overview of NAMT.} A four-plane tracker records cosmic-ray
muon hits around a sample. NAMT reads the downstream hits as residuals relative
to the upstream straight track, combines an open-field residual response with a
path-integrated scattering field, and marginalizes the unmeasured momentum
scale to obtain an event likelihood for the material map.}
\label{fig:concept}
\end{figure}

NAMT formulates the inverse problem at the hit-residual level
(Fig.~\ref{fig:concept}). The first two detector planes define the incoming
straight track, and the downstream planes provide residuals relative to that
track. A candidate $\lambda$ field predicts the sample-induced residual
variance. The unmeasured momentum is not replaced by a nominal value; NAMT
marginalizes the cosmic-ray momentum-induced scattering scale, with all
residual coordinates of the same muon sharing that scale. Non-target residuals are
learned from open-field air blanks as a conditional residual response. Object
scans then optimize only the material field. The calibration is part of the
likelihood, not a post-processing correction.

We evaluate NAMT on a controlled Geant4 benchmark spanning strong, weak, and
negative scattering contrast in dense backgrounds. The study compares PoCA,
MLS-EM, ASR, NAMT-3P, and NAMT-4P under hit-position and low-count sweeps, with
all methods scored by the same oriented top-view masks using AUC and CNR.

Our contributions are:
\begin{itemize}
\item A magnet-free muon scattering formulation that uses downstream hit
residuals directly instead of first collapsing events into PoCA points or
angle statistics.
\item A residual likelihood that combines a path-integrated radiation-length
kernel, per-muon momentum marginalization, and an open-field conditional
residual response.
\item Three-plane and four-plane NAMT variants showing that the residual
likelihood remains strong even when reconstruction uses one fewer hit plane
than the four-plane baselines.
\item A controlled benchmark across Pb, Al, RDX, and air void targets in
concrete and soil, reporting both rank separation and
background-noise-normalized contrast.
\end{itemize}

\section{Related Work}\label{sec:related}
\subsection{Statistical scattering reconstruction}
Point-of-closest-approach (PoCA) reconstruction assigns the measured bend to a
single closest-approach point~\cite{borozdin2003,schultz2004}. Angle-statistics
reconstruction (ASR) accumulates angle-based evidence over a candidate
grid~\cite{stapleton2014}. MLS-EM is closer to NAMT in spirit: it is a
model-based statistical reconstruction, using an angle--displacement datum, a
voxelized scattering-density model, and expectation-maximization
updates~\cite{schultz2007,dempster1977}. These are the three non-NAMT baselines
in our benchmark. Related statistical formulations have also treated momentum
variation as a Gaussian scale mixture~\cite{wang2009_gsm}. Relative to these
methods, NAMT changes the observation layer and the nuisance model. It places
the likelihood on downstream hit residuals, couples residual coordinates through
a shared momentum scale, and fixes the non-target residual response from
open-field data.

\subsection{Momentum and calibration}
Muon momentum information improves material discrimination because momentum is
the scale factor of multiple scattering~\cite{bonechi2020}. Prior work has
introduced additional measurements, trajectory estimators, combined scattering
with other observables, or used dedicated momentum-estimation
schemes~\cite{georgadze2024,chen2023_lowz_highz,yu2025_momentum_scheme,mutrec_2025,mutrec_app_2026,puma2025}.
NAMT addresses the no-magnet setting. It does not infer a separate momentum for
each target event; it integrates over an accepted cosmic-ray momentum prior
inside the residual likelihood. Open-field scans play the complementary role of
fixing the residual response of the instrument and environment, analogous to a
CT flat-field extended from a mean correction to a conditional distribution.

\subsection{Model-based computational imaging}
Computational imaging often keeps the physics model inside the reconstruction
and uses calibration or learned components only where the instrument response is
difficult to write analytically~\cite{kellman2019_tci,gupta2024_tci,zimmermann2024_pinqi}.
Muon imaging has also seen differentiable detector design, gradient-based
reconstruction, and learned enhancement for low-statistics
images~\cite{tomopt_2024,alameddine_2025,pezzotti2025,wang2026}. NAMT follows
the model-based route: multiple scattering and residual geometry define the
likelihood, while open-field data learn the nuisance residual distribution rather
than the material image.

\section{Method}\label{sec:method}
\subsection{Hit residuals}
Let muon $i$ have two-dimensional hits $h_{i,k}=(x_{i,k},y_{i,k})$ on detector
planes $k=1,\ldots,K$ at heights $z_1>z_2>\cdots>z_K$. NAMT uses the first two
planes to define the incident straight track. For projection
$u\in\{x,y\}$,
\begin{equation}
t_{i,u}=\frac{h_{i,2,u}-h_{i,1,u}}{z_2-z_1},
\qquad
\hat h_{i,j,u}=h_{i,1,u}+t_{i,u}(z_j-z_1),\quad j\ge3.
\end{equation}
The event datum is the downstream residual
\begin{equation}
e_{i,j,u}=h_{i,j,u}-\hat h_{i,j,u}.
\label{eq:residual}
\end{equation}
NAMT-3P uses one downstream plane, giving two residual coordinates per muon.
NAMT-4P uses two downstream planes, giving four residual coordinates. We write
the selected downstream plane indices as $\mathcal J$ and the incident condition
as $\chi_i$, the upstream information used to condition the blank response; in
the present tracker it is the incoming slope pair $(t_{i,x},t_{i,y})$.

\subsection{Material scattering kernel}
For a candidate radiation-length density field $\lambda(\mathbf r)=1/X_0$, a
small path element contributes projected angular variance
\begin{equation}
\dd\,\operatorname{Var}(\theta_u)
=g(p_i)\lambda(\mathbf r_i(\ell))\,\dd\ell,
\qquad
g(p_i)=\left(\frac{\Es}{\beta_i p_i c}\right)^2,
\label{eq:angle-scatter}
\end{equation}
where $p_i$ is the muon momentum and $\beta_i c$ is its speed. When the chord is
parameterized by detector height $z$, the path length and slope projection give
the usual $\sec^3\vartheta_i$ factor, with
$\sec\vartheta_i=\sqrt{1+t_{i,x}^2+t_{i,y}^2}$. A kick at height $z$ contributes
to downstream plane $j$ through the squared lever arm $(z_j-z)^2$. Separating
the unknown momentum scale from the geometry and material field gives the
unscaled residual-variance kernel
\begin{equation}
Q_{i,j}(\lambda)
=
\sec^3\vartheta_i
\int_{\mathrm{sample}}
(z_j-z)^2\,\lambda(\mathbf r_i(z))\,\dd z .
\label{eq:q-lambda}
\end{equation}
The material contribution to coordinate $(j,u)$ is $g(p_i)Q_{i,j}(\lambda)$.
The reported image is the depth projection of the reconstructed field.

\subsection{Open-field response and momentum marginalization}
The residuals also contain instrument and environmental structure unrelated to
the target. NAMT learns this structure from an open-field air blank. For
residual coordinate $(j,u)$ and incident condition $\chi_i$, let $b$ denote a
possible open-field residual value and let $\rho_{0,j,u}(b\mid \chi_i)$ be its
conditional density.
Given momentum scale $g$, material adds a zero-mean scattering perturbation with
variance $gQ_{i,j}(\lambda)$, so the target residual likelihood for one
coordinate is
\begin{equation}
p(e_{i,j,u}\mid\lambda,\chi_i,g)
=\int
\rho_{0,j,u}(b\mid \chi_i)
\mathcal N\!\left(e_{i,j,u}; b, gQ_{i,j}(\lambda)\right)\,\dd b.
\label{eq:coord-likelihood}
\end{equation}
This is the open-field residual density convolved with material scattering; if
$Q_{i,j}(\lambda)=0$, the coordinate likelihood reduces to the open-field
response. The same muon must share the same momentum scale across all residual
coordinates. Let $P_G$ denote the accepted distribution of the scattering scale
$g=(\Es/\beta pc)^2$ induced by the cosmic-ray momentum spectrum and detector
acceptance~\cite{gaisser1990,pdg2024}. NAMT uses a coordinate-factorized event
likelihood conditioned on this shared scale,
\begin{equation}
p(\mathbf e_i\mid\lambda,\chi_i)
=\int
\prod_{j\in\mathcal J}\prod_{u\in\{x,y\}}
p(e_{i,j,u}\mid\lambda,\chi_i,g)\,\dd P_G(g).
\label{eq:muon-likelihood}
\end{equation}

\subsection{Field reconstruction}
For $N$ accepted object muons, the object scan estimates a nonnegative voxelized
field by total-variation (TV)-regularized maximum likelihood,
\begin{equation}
\hat\lambda=
\arg\min_{\lambda\ge0}
\left[
-\frac{1}{N}\sum_i \log p(\mathbf e_i\mid \lambda,\chi_i)
 +\alpha\,\mathrm{TV}(\lambda)
\right].
\label{eq:objective}
\end{equation}
Here $\mathrm{TV}(\lambda)$ is the total variation of the voxelized
field~\cite{rof1992}. The
weight $\alpha$ is chosen before object scoring and then held fixed, as are the
open-field response, the accepted momentum-scale distribution, and the
reconstruction grid.

\FloatBarrier
\section{Experimental Setup}\label{sec:setup}
\subsection{Benchmark scenes}
\begin{figure}[!htbp]
\centering
\includegraphics[width=0.70\linewidth]{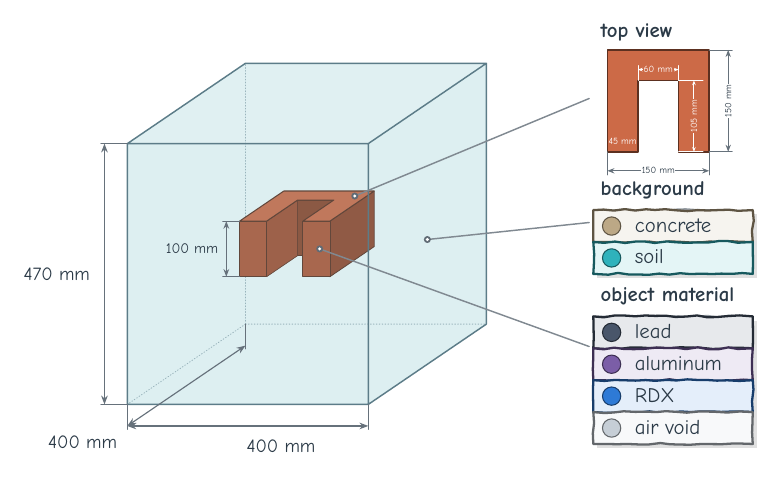}
\caption{\textbf{Benchmark geometry.} Each scene embeds a centered U-shaped
block in a $400\times400\times470$~mm background. The eight object scenes pair
four materials (lead, aluminum, RDX, and air void) with two backgrounds
(concrete and soil).}
\label{fig:setup}
\end{figure}

The benchmark uses Geant4 transport~\cite{geant4}, CRY cosmic-ray muon
generation~\cite{cry}, and RPC tracking planes~\cite{santonico1981}. Each scene
contains one centered U-shaped object in a
$400\times400\times470$~mm background volume (Fig.~\ref{fig:setup}). The
geometry keeps the shape fixed while changing the material contrast: Pb tests
strong positive scattering contrast, aluminum tests weak positive contrast, RDX
tests near-background weak contrast, and an air void tests negative contrast. Such
weak-scattering and negative-contrast objects are established stress cases for
muon security imaging~\cite{cuellar2009_explosives,reed2025_lowz}. The two
backgrounds are concrete and soil.

\begin{table}[!ht]
\centering
\caption{Benchmark materials. The electron-density proxy $s=\rho Z/A$ and
radiation length $X_0$ define the simulated scenes.}
\label{tab:materials}
\setlength{\tabcolsep}{8pt}
\renewcommand{\arraystretch}{1.20}\footnotesize
\begin{tabular}{@{}lcc@{}}
\toprule
material & $s=\rho Z/A$ & $X_0$ (mm) \\
\midrule
\multicolumn{3}{@{}l}{\textbf{Background}}\\
\quad concrete & 1.171 & 115.53 \\
\quad soil & 0.8529 & 159.60 \\
\addlinespace[2pt]
\multicolumn{3}{@{}l}{\textbf{Object}}\\
\quad lead & 4.492 & 5.6125 \\
\quad aluminum & 1.301 & 88.97 \\
\quad RDX & 0.9536 & 200.73 \\
\quad air void & 0.0015 & $3.0392\times10^5$ \\
\bottomrule
\end{tabular}
\end{table}

\subsection{Compared methods and protocol}
All methods are run on the same accepted event set and scored on the same
top-view grid. The benchmark includes PoCA, MLS-EM, ASR, NAMT-3P, and NAMT-4P.
PoCA, ASR, and MLS-EM use four-plane upstream--downstream track pairs. NAMT-3P
uses one downstream residual plane, and NAMT-4P uses two, so the 3P/4P
comparison isolates the value of the additional downstream hit while keeping the
event set fixed.

The reference setting uses $120$k effective muons and
$\sigpos=1$~mm hit-position error. We then vary hit error
($\sigpos\in\{1,2,3\}$~mm at $120$k muons) and exposure ($120$k, $60$k, and
$30$k effective muons at $\sigpos=1$~mm). Target masks are used only for
evaluation. Here, effective muons are accepted events entering reconstruction
after the common event selection.

\subsection{Metrics}
The primary score is area under the ROC curve (AUC). Each released map is
multiplied by a contrast sign fixed from the known target--background
radiation-length contrast before metric computation, so target pixels are
expected to be larger than background pixels. Masks are then used only to sample
target and background pixels. AUC is
\begin{equation}
\mathrm{AUC}
=\Pr(A_{\mathrm{target}}>A_{\mathrm{bg}})
+\frac{1}{2}\Pr(A_{\mathrm{target}}=A_{\mathrm{bg}}).
\end{equation}
Here $A_{\mathrm{target}}$ and $A_{\mathrm{bg}}$ are independently sampled
oriented map values from the target and background masks.
We also report contrast-to-noise ratio (CNR) to quantify amplitude separation
in background-noise units,
\begin{equation}
\mathrm{CNR}=
\frac{\bar v_t-\bar v_b}{\operatorname{std}(v_b)},
\label{eq:cnr}
\end{equation}
where $v_t$ and $v_b$ are oriented target and background values. AUC measures
rank separation between target and background pixels; CNR measures how far the
target mean sits above the background fluctuations in the oriented image.

\FloatBarrier
\section{Results}\label{sec:results}
\subsection{Reference performance}
Table~\ref{tab:ref-auc} reports per-scene AUC at the reference setting. NAMT-4P
has the highest mean AUC, $0.916$, and NAMT-3P remains close at $0.909$ despite
using only one downstream residual plane. Both exceed the strongest non-NAMT
mean, ASR ($0.784$), and the model-based MLS-EM baseline ($0.749$). The Pb
scenes are nearly saturated for MLS-EM, ASR, and NAMT; the method separation
appears in lower-contrast and negative-contrast scenes, where nuisance variance
can otherwise obscure the material signal. For example, in Al/concrete, NAMT-4P
reaches $0.923$ versus $0.548$ for ASR and $0.529$ for MLS-EM; in RDX/soil,
NAMT-3P reaches $0.842$ and NAMT-4P reaches $0.817$, compared with $0.610$ for
ASR and $0.640$ for MLS-EM.

\begin{table}[H]
\centering
\caption{\textbf{Per-scene AUC} at $120$k effective muons and
$\sigpos=1$~mm. Columns group object material with concrete and soil
backgrounds. Bold marks the best value in each column.}
\label{tab:ref-auc}
\setlength{\tabcolsep}{3.2pt}
\renewcommand{\arraystretch}{1.22}\footnotesize
\begin{tabular}{@{}l*{9}{c}@{}}
\toprule
& \multicolumn{2}{c}{Pb} & \multicolumn{2}{c}{Al} & \multicolumn{2}{c}{RDX} & \multicolumn{2}{c}{void} & \\
\cmidrule(lr){2-3}\cmidrule(lr){4-5}\cmidrule(lr){6-7}\cmidrule(lr){8-9}
method & conc. & soil & conc. & soil & conc. & soil & conc. & soil & mean \\
\midrule
\quad PoCA & 0.926 & 0.938 & 0.566 & 0.648 & 0.514 & 0.451 & 0.568 & 0.536 & 0.643 \\
\quad MLS-EM & 0.990 & 0.992 & 0.529 & 0.646 & 0.647 & 0.640 & 0.800 & 0.749 & 0.749 \\
\quad ASR & 0.992 & 0.994 & 0.548 & 0.706 & 0.723 & 0.610 & 0.892 & 0.804 & 0.784 \\
\quad NAMT-3P & \best{1.000} & \best{1.000} & 0.921 & 0.675 & \best{0.925} & \best{0.842} & 0.966 & 0.947 & 0.909 \\
\quad NAMT-4P & 0.999 & \best{1.000} & \best{0.923} & \best{0.744} & 0.923 & 0.817 & \best{0.972} & \best{0.949} & \best{0.916} \\
\bottomrule
\end{tabular}
\end{table}

CNR gives the amplitude view of the same separation
(Table~\ref{tab:ref-cnr}). NAMT-3P has the highest mean CNR, $3.68$, followed
by NAMT-4P at $3.20$; the strongest non-NAMT mean is MLS-EM at $1.80$.
Together, AUC and CNR expose the 3P/4P tradeoff: four planes give the best mean
rank separation, whereas the three-plane variant often gives stronger
target--background amplitude contrast.

\begin{table}[H]
\centering
\caption{\textbf{Per-scene CNR} at $120$k effective muons and
$\sigpos=1$~mm. Values are target--background mean separation in background
standard-deviation units after orientation. Bold marks the best value in each
column.}
\label{tab:ref-cnr}
\setlength{\tabcolsep}{3.2pt}
\renewcommand{\arraystretch}{1.22}\footnotesize
\begin{tabular}{@{}l*{9}{c}@{}}
\toprule
& \multicolumn{2}{c}{Pb} & \multicolumn{2}{c}{Al} & \multicolumn{2}{c}{RDX} & \multicolumn{2}{c}{void} & \\
\cmidrule(lr){2-3}\cmidrule(lr){4-5}\cmidrule(lr){6-7}\cmidrule(lr){8-9}
method & conc. & soil & conc. & soil & conc. & soil & conc. & soil & mean \\
\midrule
\quad PoCA & 2.00 & 1.88 & 0.04 & 0.21 & 0.13 & 0.03 & 0.17 & 0.15 & 0.58 \\
\quad MLS-EM & 5.66 & 6.29 & -0.07 & 0.31 & 0.44 & 0.42 & 0.72 & 0.68 & 1.80 \\
\quad ASR & 3.40 & 3.46 & 0.08 & 0.58 & 0.67 & 0.34 & 1.19 & 0.97 & 1.34 \\
\quad NAMT-3P & \best{8.41} & \best{9.27} & \best{1.32} & 0.66 & \best{2.24} & \best{1.27} & \best{3.36} & \best{2.87} & \best{3.68} \\
\quad NAMT-4P & 6.85 & 7.88 & 1.31 & \best{0.91} & 1.98 & 1.14 & 3.12 & 2.43 & 3.20 \\
\bottomrule
\end{tabular}
\end{table}

\subsection{Robustness to hit error and low counts}
Tables~\ref{tab:robust} and~\ref{tab:robust-cnr} summarize eight-scene mean AUC
and CNR under the two sweeps. NAMT-4P is the most stable method as hit error
increases from $1$ to $3$~mm, remaining at $0.907$ mean AUC at the largest
error. CNR sharpens the 3P/4P tradeoff: NAMT-3P has the highest mean CNR at
$1$ and $2$~mm, while NAMT-4P is slightly higher at $3$~mm. At the lowest
exposure, the three-plane likelihood becomes the strongest variant in both
metrics. At $30$k effective muons, NAMT-3P reaches $0.804$ mean AUC and $2.37$
mean CNR, above NAMT-4P ($0.783$, $2.09$) and all three baselines. Thus the
additional fourth plane mainly improves hit-error robustness; it is not
required for NAMT to outperform the four-plane baselines.

\begin{table}[H]
\centering
\caption{\textbf{All-scene robustness}: eight-scene mean AUC under larger
hit-position error and reduced exposure. Bold marks the best
value in each column.}
\label{tab:robust}
\setlength{\tabcolsep}{4pt}
\renewcommand{\arraystretch}{1.22}\footnotesize
\begin{tabular}{@{}lccc@{\quad}c@{\quad}ccc@{}}
\toprule
& \multicolumn{3}{c}{position error $\sigpos$ (mm)} & & \multicolumn{3}{c}{effective muons} \\
\cmidrule(lr){2-4}\cmidrule(l){6-8}
method & 1 & 2 & 3 & & 120k & 60k & 30k \\
\midrule
\quad PoCA & 0.643 & 0.634 & 0.625 & & 0.643 & 0.623 & 0.617 \\
\quad MLS-EM & 0.749 & 0.717 & 0.696 & & 0.749 & 0.710 & 0.665 \\
\quad ASR & 0.784 & 0.763 & 0.739 & & 0.784 & 0.748 & 0.708 \\
\quad NAMT-3P & 0.909 & 0.894 & 0.867 & & 0.909 & 0.854 & \best{0.804} \\
\quad NAMT-4P & \best{0.916} & \best{0.910} & \best{0.907} & & \best{0.916} & \best{0.863} & 0.783 \\
\bottomrule
\end{tabular}
\end{table}

\begin{table}[H]
\centering
\caption{\textbf{All-scene robustness in CNR}: eight-scene mean CNR under
larger hit-position error and reduced exposure. Bold marks the best value in
each column.}
\label{tab:robust-cnr}
\setlength{\tabcolsep}{4pt}
\renewcommand{\arraystretch}{1.22}\footnotesize
\begin{tabular}{@{}lccc@{\quad}c@{\quad}ccc@{}}
\toprule
& \multicolumn{3}{c}{position error $\sigpos$ (mm)} & & \multicolumn{3}{c}{effective muons} \\
\cmidrule(lr){2-4}\cmidrule(l){6-8}
method & 1 & 2 & 3 & & 120k & 60k & 30k \\
\midrule
\quad PoCA & 0.58 & 0.54 & 0.53 & & 0.58 & 0.47 & 0.36 \\
\quad MLS-EM & 1.80 & 1.22 & 0.91 & & 1.80 & 1.60 & 1.05 \\
\quad ASR & 1.34 & 1.21 & 1.09 & & 1.34 & 1.15 & 0.94 \\
\quad NAMT-3P & \best{3.68} & \best{2.83} & 2.32 & & \best{3.68} & \best{3.19} & \best{2.37} \\
\quad NAMT-4P & 3.20 & 2.69 & \best{2.38} & & 3.20 & 2.81 & 2.09 \\
\bottomrule
\end{tabular}
\end{table}

\FloatBarrier
\section{Discussion and Limitations}\label{sec:discussion}
NAMT's advantage is best understood relative to MLS-EM. Both are model-based
statistical reconstructions of a voxelized scattering field, but MLS-EM works
with an angle--displacement datum formed from upstream and downstream track
segments. The gain over MLS-EM suggests that the decisive change is not simply
using a likelihood; it is placing that likelihood at the downstream-hit residual
layer and calibrating non-target residuals before fitting material. The method
separation is therefore most visible outside saturated Pb scenes, where target
contrast must be recovered in the presence of momentum and open-field nuisance
variation.

The three-plane result is also part of the method, not only an ablation.
NAMT-3P reconstructs from one downstream plane after the two upstream planes,
whereas PoCA, ASR, and MLS-EM use four-plane track pairs. Its reference AUC is
nearly the same as NAMT-4P, its reference CNR is the highest in the benchmark,
and it is the best method at the lowest exposure. NAMT-4P is still more stable
under larger hit-position errors because the additional downstream plane adds
residual coordinates and geometric leverage; both variants are evaluated on the
same accepted event set.

\begin{figure}[!t]
\centering
\includegraphics[width=\linewidth]{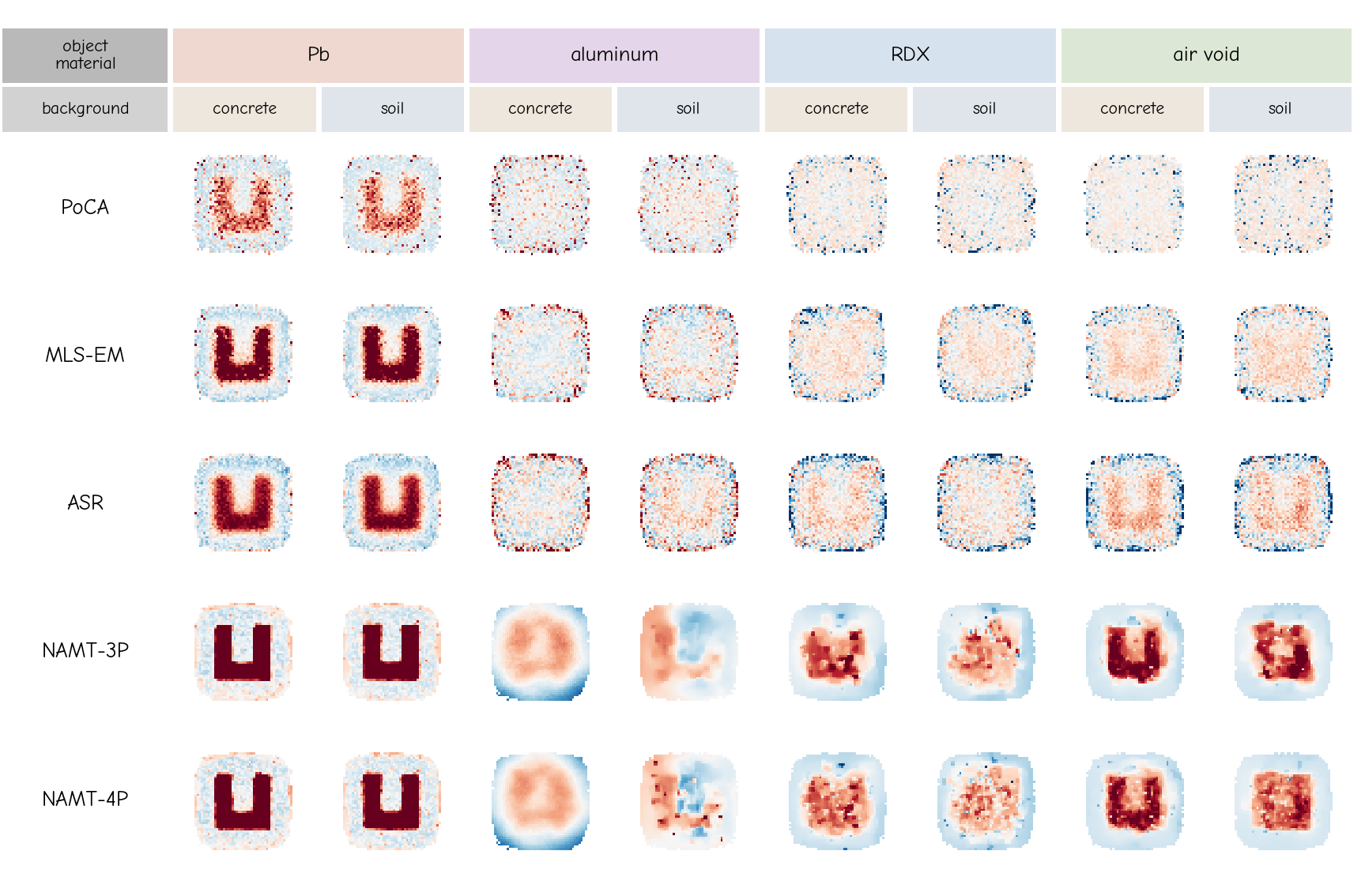}
\caption{\textbf{Reconstruction gallery} at $120$k effective muons and
$\sigpos=1$~mm. Rows show PoCA, MLS-EM, ASR, NAMT-3P, and NAMT-4P. Columns show
Pb, aluminum, RDX, and air void targets, each in concrete and soil. Maps are
oriented so the target has positive contrast and each panel is standardized by
its own background standard deviation; red indicates the target direction. The
panels show presence contrast and background control rather than high-detail
shape recovery.}
\label{fig:gallery}
\end{figure}

Figure~\ref{fig:gallery} provides the qualitative counterpart to these
quantitative results. NAMT gives cleaner object support in Pb and several void
scenes. In aluminum and RDX scenes embedded in concrete or soil, the main gain
is improved presence recognition; fine shape remains blurred or fragmented.

These aluminum and RDX scenes also mark the present boundary. The best NAMT
reconstruction improves AUC and CNR over the baselines and is therefore better
at identifying that an object is present. Because event momentum is still
unmeasured, however, fine geometry and material discrimination remain difficult
in such near-background settings. Similar weak-contrast security scenarios will
likely require event-level momentum, or an equivalent additional constraint,
when detailed discrimination is required.

\section{Conclusion}\label{sec:conclusion}
NAMT casts magnet-free cosmic-ray muon scattering imaging as an
open-field-calibrated residual-likelihood inverse problem. It uses the first
two hits to define the
incoming track, models downstream residual variance with a path integral of
$\lambda=1/X_0$, marginalizes the unknown momentum scale over the accepted
cosmic-ray prior, and learns non-target residuals from air blanks. On eight Pb,
Al, RDX, and air-void benchmark scenes, the NAMT variants give the strongest
mean AUC and CNR at the reference setting, with NAMT-4P leading mean AUC and
NAMT-3P leading mean CNR. Four-plane NAMT gives the best hit-error robustness.
Three-plane NAMT preserves most of the AUC while using one fewer hit plane in
reconstruction than the baselines, and is strongest in the lowest-count setting.

\section*{Acknowledgment}
During the preparation of this manuscript, the authors used AI writing
assistants for language editing, stylistic refinement, and drafting assistance
from author-provided outlines, results, and technical direction. The authors
reviewed, edited, and verified all resulting text and take full responsibility
for the content. All scientific ideas, methodology, experiments, figures,
tables, and conclusions are the authors' own.

\bibliographystyle{plain}
\bibliography{refs/refs_muon,refs/newrefs}

\appendix
\begin{center}{\Large\bfseries\scshape Appendix}\end{center}
\vspace{0.4em}
%================================================================
\section{Residual-kernel details}\label{app:residual-kernel}
For an accepted muon, the upstream hits define an incident chord
\[
\mathbf r_i(z)=
\bigl(h_{i,1,x}+t_{i,x}(z-z_1),\,
h_{i,1,y}+t_{i,y}(z-z_1),\,z\bigr),
\qquad
\sec\vartheta_i=\sqrt{1+t_{i,x}^2+t_{i,y}^2}.
\]
The height coordinate is convenient because detector planes are indexed by
their heights. A path element satisfies $\dd\ell=\sec\vartheta_i\,\dd z$.
If the projected slope kick at height $z$ is $\dd t_u(z)$, then its contribution
to downstream plane $j$ is
\[
\dd h_{j,u}=(z_j-z)\,\dd t_u(z),
\]
up to the sign convention used for slopes and detector height. The variance
calculation depends on the squared lever arm, so the sign drops out.

Using the small-angle Gaussian core of multiple Coulomb scattering, a material
element contributes
\[
\dd\,\operatorname{Var}(t_u)
=
\left(\frac{\Es}{\beta p c}\right)^2
\lambda(\mathbf r_i(z))\sec^3\vartheta_i\,\dd z .
\]
The first factor is the momentum-dependent scattering scale. The factor
$\sec\vartheta_i$ comes from the longer path through the material, and the
remaining projection factors come from expressing the kick as a slope with
respect to detector height. Propagating the independent variance increments to
plane $j$ gives
\[
\operatorname{Var}(e^{\mathrm{mat}}_{i,j,u}\mid p_i,\lambda)
=
g(p_i)\,
\sec^3\vartheta_i
\int_{\mathrm{sample}}
(z_j-z)^2\lambda(\mathbf r_i(z))\,\dd z
=g(p_i)Q_{i,j}(\lambda).
\]
This is Eq.~\eqref{eq:q-lambda} in variance form. On a voxel grid with constant
value $\lambda_v$ in voxel $V_v$, the same kernel is linear:
\[
Q_{i,j}(\lambda)=\sum_v W_{i,jv}\lambda_v,\qquad
W_{i,jv}=\sec^3\vartheta_i
\int_{\mathrm{sample}\cap V_v}
(z_j-z)^2\,\dd z .
\]
Thus the chord selects which voxels can contribute, and the squared lever arm
sets how strongly each local scattering variance appears in the downstream
residual.

\section{Open-field response model}\label{app:blank-response}
The air blank is processed with the same hit selection, hit smearing, and
incident-track construction as object scans. Since the blank contains no target,
its residuals estimate the non-target response of the instrument and
environment. For one residual coordinate, NAMT writes the object-scan residual
as
\[
E_{i,j,u}=B_{i,j,u}+S_{i,j,u},
\]
where $B_{i,j,u}\mid\chi_i$ has the fixed open-field density
$\rho_{0,j,u}(\cdot\mid\chi_i)$, and the material term satisfies
\[
S_{i,j,u}\mid g,\lambda \sim
\mathcal N\!\left(0,\,gQ_{i,j}(\lambda)\right).
\]
Conditioned on the momentum scale $g$, this gives the convolution likelihood
\[
p(e_{i,j,u}\mid\lambda,\chi_i,g)
=
\int
\rho_{0,j,u}(b\mid\chi_i)
\mathcal N\!\left(e_{i,j,u};b,gQ_{i,j}(\lambda)\right)\,\dd b .
\]
The null-target case is recovered by setting $Q_{i,j}(\lambda)=0$, in which
case the likelihood reduces to the open-field response.

The construction is not tied to a particular density family. For example, if
the blank response is represented as a Gaussian mixture, convolution simply
adds the material variance to each component,
\[
\left[
\mathcal N(\cdot;\mu_m,\tau_m^2)
\ast
\mathcal N(\cdot;0,gQ_{i,j}(\lambda))
\right](e)
=
\mathcal N\!\left(e;\mu_m,\tau_m^2+gQ_{i,j}(\lambda)\right),
\]
with the mixture weights inherited from the blank response. The likelihood
therefore requires a calibrated conditional density, not a specific estimator.

The momentum scale is shared by all residual coordinates of the same muon.
Equivalently, with $G\sim P_G$,
\[
p(\mathbf e_i\mid\lambda,\chi_i)
=
\mathbb E_G
\left[
\prod_{j\in\mathcal J}\prod_{u\in\{x,y\}}
p(e_{i,j,u}\mid\lambda,\chi_i,G)
\right].
\]
This shared integration is the part that prevents a single large residual
coordinate from being interpreted independently of the rest of the event.

\section{Compared methods}\label{app:baselines}
PoCA, ASR, and MLS-EM use four detector planes to form upstream and downstream
track segments. Let $\mathbf r_i^{\mathrm{up}}(z)$ and
$\mathbf r_i^{\mathrm{dn}}(z)$ denote these two fitted segments, and let
$\boldsymbol\theta_i^{\mathrm{up}}$ and
$\boldsymbol\theta_i^{\mathrm{dn}}$ be their projected angles. The measured
bend score is
\[
a_i=
\left\|
\boldsymbol\theta_i^{\mathrm{dn}}-
\boldsymbol\theta_i^{\mathrm{up}}
\right\|^2 .
\]

PoCA assigns this bend to a single estimated scattering point. The implemented
single-depth approximation uses the coordinate-wise crossing depths
$z_{i,u}^{\mathrm{cross}}$ defined by
$r_{i,u}^{\mathrm{up}}(z_{i,u}^{\mathrm{cross}})
=r_{i,u}^{\mathrm{dn}}(z_{i,u}^{\mathrm{cross}})$ and writes
\[
z_i^\star=
\Pi_{\mathrm{sample}}\!\left(
\frac{1}{2}\sum_{u\in\{x,y\}}z_{i,u}^{\mathrm{cross}}
\right),
\qquad
\mathbf c_i=\mathbf r_i^{\mathrm{up}}(z_i^\star),
\]
where $\Pi_{\mathrm{sample}}$ clips the depth to the sample interval, with the
mid-height fallback used for nearly parallel projections. The top-view PoCA
image averages $a_i$ over events whose projected PoCA point falls in a grid
cell.

ASR keeps the angle statistic but avoids committing the event to a single
closest-approach depth. For a set of candidate depths $\{z_k\}$ through the
sample, it back-projects the absolute projected bend along the incident chord.
For a top-view cell $c$, the score has the form
\[
S_{\mathrm{ASR}}(c)=
\frac{1}{|\mathcal I_c|}
\sum_{(i,k,u)\in\mathcal I_c}
\left|
\theta^{\mathrm{dn}}_{i,u}-\theta^{\mathrm{up}}_{i,u}
\right|,
\]
where $\mathcal I_c$ contains the event-depth-coordinate contributions whose
projected chord positions fall in cell $c$.

MLS-EM is closer in spirit to NAMT because it is a statistical inverse method.
For each projection $u$, it forms an angle--displacement datum
\[
\mathbf y_{i,u}=
\begin{bmatrix}
\Delta\theta_{i,u}\\
\Delta x_{i,u}
\end{bmatrix},
\]
and models it with a covariance built from a voxelized scattering field:
\[
\Sigma_i(\lambda)=\Sigma_{\mathrm{hit},i}+\sum_v \lambda_v M_{iv}.
\]
Here $\Sigma_{\mathrm{hit},i}$ is the hit-resolution contribution and $M_{iv}$
is the geometric moment matrix for voxel $v$ along event $i$. The Gaussian
negative log-likelihood is
\[
\mathcal L_{\mathrm{MLS}}(\lambda)=
\sum_{i,u}
\left[
\log\det\Sigma_i(\lambda)
+\mathbf y_{i,u}^{\mathsf T}
\Sigma_i(\lambda)^{-1}
\mathbf y_{i,u}
\right],
\]
and the image is obtained by EM-style updates of the scattering field. This
keeps a physical covariance model, but it operates after the hits have already
been compressed into upstream--downstream angle and displacement observables.

NAMT-3P and NAMT-4P differ only in the residual coordinates used for
reconstruction: NAMT-3P uses one downstream residual plane after the two
upstream planes, whereas NAMT-4P uses both downstream residual planes.

\end{document}